# Chapter IX

# E-Business Experiences with Online Auctions

Bernhard Rumpe
Munich University of Technology, Germany

## ABSTRACT

*Online auctions are among the most influential e-business applications. Their impact on trading for businesses, as well as consumers, is both remarkable and inevitable. There have been considerable efforts in setting up market places, but, with respects to market volume, online trading is still in its early stages. This chapter discusses the benefits of the concept of Internet marketplaces, with the highest impact on pricing strategies, namely, the conduction of online business auctions. We discuss their benefits, problems and possible solutions. In addition, we sketch actions for suppliers to achieve a better strategic position in the upcoming Internet market places.*

## INTRODUCTION

Electronic commerce will be the enabling technology for the forthcoming revolution in local and global trading. Virtual, Internet-based markets allow for entirely different forms of trading (Höller et. al., 1998) than are known so far. Local and, therefore, sometimes monopolistic markets become global and more competitive.

Expectations are that, in the forthcoming decade, the Internet will be a market-enabler of unforeseen possibilities. Just in the last few years, it became apparent that doing business on the Internet can simplify marketing and purchasing considerably. Figure 1 shows estimates of the worldwide trading volume on the Internet (Forrester Research, 1999) that remarkably still hold. The part of e-business that is based on online auctions is growing equally. Therefore, an increasing number of online marketplaces have come into existence. Online marketplaces simplify the establishment of new alternative purchasing and selling partnerships. In the C2C and B2C areas, this fact has been widely recognized, and numerous new B2B marketplaces have been emerging in the last two years.

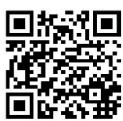

In most market places, sellers may advertise their offers, and consumers and industrial purchasers can distribute their demands via the Internet. Whereas these forms of establishing connections are important, it is the use of online auction systems that has an effective impact on the pricing structure. Auctioning is among the most efficient and fastest concepts available to achieve fair and competitive prices and identify the optimal business partner.

The chart in Figure 2 shows the core issues of electronic sourcing, which can be separated in an economic and a technical layer. The technical layer includes exchangeable documents based on the EDI standard, e.g., EDIFACT (ISO, 1993) or (UN/EDIFACT, 1993), or XML as a technical infrastructure (W3 Consortium, 2000); protocols for their safe transmission; and electronic payment systems, etc. The technical layer strongly supports, and is driven by, the economic layer. The economic layer focuses on the introduction of new strategies and techniques to let the vision of the Internet economy come true. This includes new interfaces to other companies, as well as restructuring of the company internal business processes, with the goal of overall improvement and cost reduction. The chart also shows that, among the existing concepts, the potentials and effects, together with the impact on the organization and, most importantly, the potential gain, grows at its best when using Internet auctions. Whereas the other concepts mainly concentrate on improving the overall sales and purchasing process, the use of Internet auctions has an impact on the pricing structure of traded goods.

*Figure 1: Worldwide trading volume on the Internet (Forrester Research, 1999)*

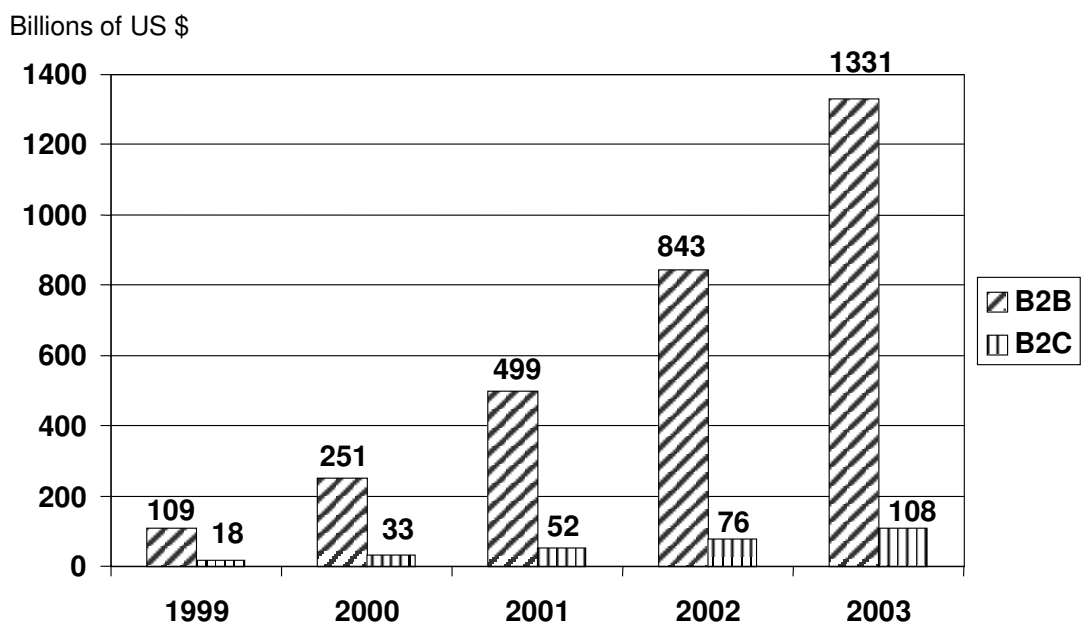

*Figure 2: Electronic sourcing*

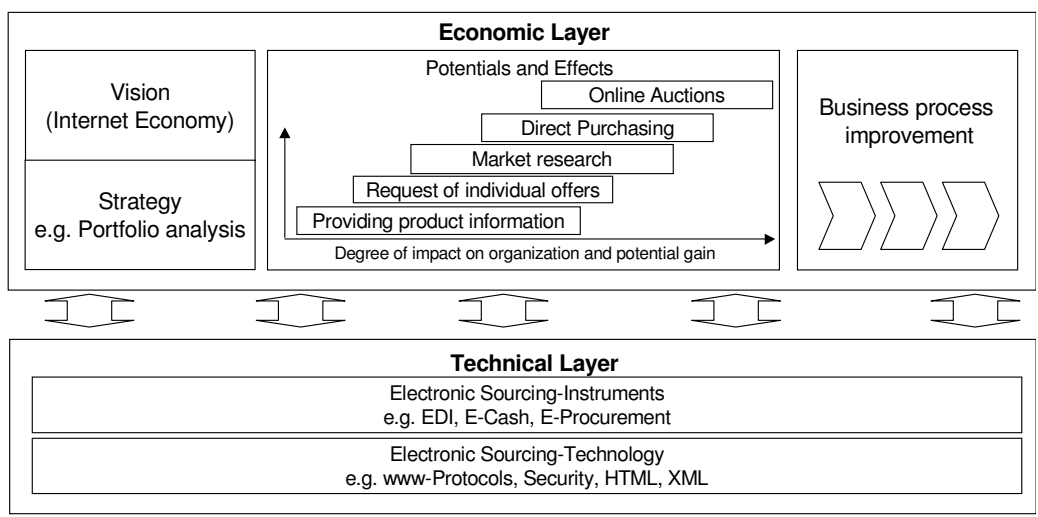

Many marketplaces focus on e-procurement elements when trying to make the purchasing process more efficient. Contact with professional purchasing departments of large and middle-sized European companies shows that, by far, the largest gain in time and money arises from a proper identification of materials that qualify for Internet auctions, a set of possible suppliers and an appropriate auction format.

# AUCTION FORMATS

The real world already has defined the primary auction formats that are possible and useful. Standard (also called English) auctions are used to offer one product to several potential buyers. On the contrary, reverse auctions (also called purchaser's auctions) allow one purchaser to auction a demand among several potential suppliers. In reverse auctions, the bidders submit decreasing bids.

If a supplier has more than one item to sell (e.g., a fixed number of tulip bulbs), he may use the Dutch auction format. The price for one product constantly decreases during a period of time. Whenever a bidder submits a bid, he gets one item for the valid price at that time. The auction is finished when all items are sold.

The use of electronic resources invites bidders throughout the world to participate in the auction simultaneously and provides for more complex and, depending on the particular situation, more specific auction formats. Long-term auctions may last up to four weeks, throughout which the bidders repeatedly review the current situation. However, it appears that such auctions are of interest mainly during the closing phase. Short-term auctions concentrate on these last few hours immediately. They may even be as short as 30 minutes, provided that the participants are invited beforehand.

Further, auction formats, such as multi-round biddings, have been defined. In a multi-round bidding, each bidder is forced to submit exactly one bid per round. All bidders are informed of their competitor's bids only after the round has ended, and the next round begins. Multi-phase auctions are quite similar; After each phase, only a subset of the earlier bidders is admitted to proceed to the next phase.

Recently, experience has been gained in conducting multi-dimensional auctions. Here, the price is left open and negotiated through the auction, and several other variables are determined, as well. For example, the price may be combined from the supplier and the logistics entrepreneur (Prince, 1999).

In running Internet auctions, it became increasingly apparent that the auctioneer must be independent to ensure both supplier and buyer have enough confidence in the fairness of the auction process. The experiences with other marketplaces have shown, in an apparent way, that an auction marketplace is not very successful when operated by the buyers or sellers themselves.

Auctions also can be much more complicated. An auction can, for example, exist of multiple slots that allow the slicing of materials and goods in order to auction them among several suppliers. This setting is important to prevent dependency from a single supplier – a critical issue when strategic goods are involved. These slots also may depend on each other. For example, if two competing types of materials can replace each other, and the price difference between both determines the purchased quantities. Another interesting question focuses on determining at what time certain information should be revealed to specific participants. This should prevent illegal price agreements between bidders and, at the same time, ensure confidence in a fair auction. For example, should competitors know each other? An interesting variant reveals identities of bidders at the beginning of a very short-term auction.

These examples show that, today, the power of Internet auctions has been by no means explored. Moreover, there are still a number of unforeseen variations to come.

Having conducted quite a number of online auctions, it became obvious that identifying auctionable goods and materials is not an easy task. Therefore, it is a common way to rely on the assistance of a consultant to define the actual auction set-up, starting with the identification of the demands and possible suppliers. On the other hand, it also turned out that it is rather irrelevant to have large supplier lists at hand, because companies that buy material in industrial sizes usually know their probable suppliers beforehand. They keep watching the suppliers' situation and they want certifications that prove the suppliers' capability of delivering high quality material in time.

# FEATURES OF AN ONLINE AUCTION SYSTEM

In several seminars with major industrial purchasers, it soon became apparent that complete e-Procurement strategies are a time-consuming task to define and implement. On the one hand, aligning a whole purchasing department with electronic procurement is tiresome and brings an extra load of work. On the other hand, the desired effects of more efficient purchasing processes pay off only in the long run. However, it also became apparent that identifying the commodities with high purchase volume and buying them via auctions could significantly reduce the prices spent for these commodities and, what is most important, leads to a high and immediate return on investment. From that point of view, an initial set of requirements for online auctions was defined: It was evolved and enhanced during the following period. In the following, the most important of these features have been described. Also, other technical issues of interest, such as flexibility, ease of use, security, performance, robustness, and compatibility to existing systems, both on the buyer's as well as the supplier's side have been discussed.

1. The software is capable of online auctions, both the normal English format and the reverse format, thus allowing the auction of goods for buyers and demands for suppliers.
2. An intuitive graphical user interface is accessible through the Web, without any installation necessary. Being able to support access to online auctions without any installation proved a great success factor, because it often is difficult to install new software within company networks.
3. Auctions are running in real-time. This means that clients always have current information visible. This is especially important for short time (approximately 30- minute) auctions, where the frequency of bids is relatively high.
4. An auction may consist of several slots, allowing the buyer to split the material desired among several suppliers. This prevents dependency on a single supplier, and enables the auction to split the material for different delivery points. Later, we will discuss a mechanism to extend the auction time automatically, in order to ensure competitors are able to react to incoming bids. However, this extension time will get shorter as the auction proceeds. We had auctions where the bids arrived within seconds.
5. Different auctions may depend on each other. For example, depending on the results of simultaneous auctions, the buyer purchases percentages of competing materials. The auction system must reflect this dependency, e.g., with additional messages that describe which of the competing materials will be bought.

6. Persons may participate in an auction in different roles. The auctioneer, the bidders, the originator of the auction (buyer in reverse auctions, seller in the normal auctions) and guests shall be admitted. In particular, the role of guest is useful to show potential — and not fully convinced — participants how online auctions work, without revealing any information on auction details (neither currency, nor value, nor the buyer, nor the kind of traded goods).
7. Different roles receive different available information. Only the auctioneer can co-relate the bids to their bidders during the auction. Bidders appear to each other anonymously, but know how many competitors exist. Furthermore, bidders see their ranking. External observers following the auction see percentage values, instead of real currency.
8. Reverse auctions may have a historic and a target value. The historic value describes what the buyer paid for the auction goods so far, whereas the target value describes what he would like to pay this time. If the auction result hits the target value, then the buyer is obliged to sign the contract. If the target prize is not hit, the buyer is free to choose.
9. The auction times may vary. Very short auctions may have an auction time as short as 15 minutes. Typical auction times are between one and three hours, consisting of a main part and an extension part.
10. The auction time is extended whenever a bid arrives shortly before the auction's end. This allows all other bidders to react. The provided reaction time may vary, for example, starting from three minutes as an initial extension down to a few seconds at the very end.
11. A login mechanism is imperative. Passwords are distributed through safe channels, among them PGP-encrypted e-mails.
12. A report on the auction results is provided for all participants. This report allows the participants to view the auction results. The winner has evidence of his success. The other bidders have evidence that they have been out-bidden and perhaps should think about the pricing structure of their products.

Although the field of e-commerce is highly innovative, to our current knowledge (August, 2000), no existing online auction system provides all the functionality characterized above (Glänzer & Schäfers, 2000), (Grebe & Samwer, 2000), (Wahrenberg, 2000). Therefore, we created our own auction system, which is now online since March 2000. After two phases of intensive elaboration, a strongly increasing number of online auctions has been conducted. The rest of this paper discusses the lessons learned from these online auctions, with this system, for major industrial companies in Europe.

The Emporias' auction engine systems can cope with the above described auction formats. It provides standard and extension phases, historical and target prices, multiple slots, visibility constraints for multiple participants (bidder, guests with different access rights), and a ranking of bidders and other participants. Based on the

*Figure 3: Screenshot of an Emporias online auction*

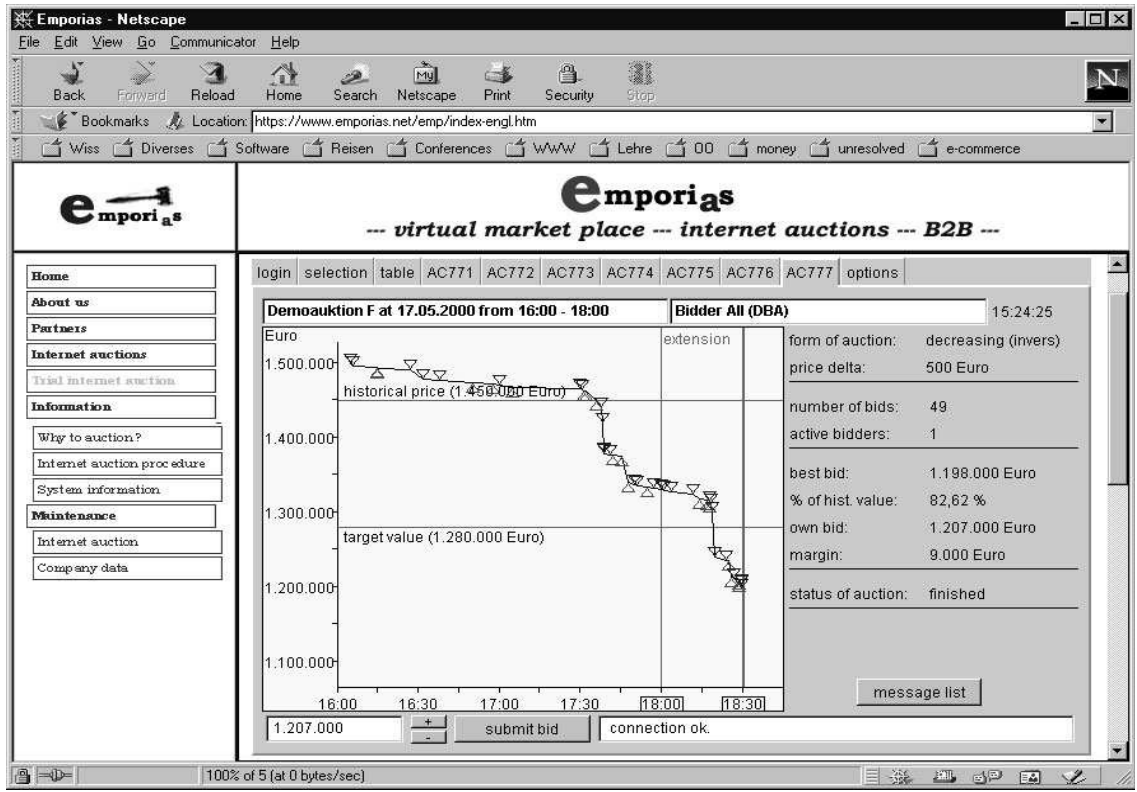

Java technology used for implementation and, therefore, its emerging Internet-realtime features, it is capable of active monitoring of short-time auctions, as well as bidding agents.

# LESSONS LEARNED

So far, we have conducted online auctions for a volume of almost $100 million (U.S.) These auctions have occurred in different formats, settings and, of course, different market types. In the following, we concentrate on the results from reverse auctions, because most of the conducted auctions were this type. We have learned a lot about the differentiation between various markets, problems that can be encountered and the solutions to master these problems. Therefore, we concentrate on the most common problems encountered and sketch solutions for them.

## Emotional refusal of the Internet

Although — in the C2C and the B2B market — the number of online customers is increasing exponentially, knowledge of the Internet in the area of B2B purchasing is still somewhat limited. In particular, many managers do not trust the Internet's security and ability to conduct safe trading online. That emotional barrier can be reduced only by demonstrating – to such a manager — how an online auction works. For that purpose, the guest role for online auctions has been implemented, enabling us to invite foreign guests to participate in an auction, without giving them knowledge of auction details (e.g., price). Furthermore, it helps if, at least for the first auctions, a consultant is assisting the setup process.

## Political resistance

As always, when people collaborate, different interests may give rise to conflict. In particular, some people are interested in improving the purchasing process, whereas others would like to stay with the old, approved paths, because they do not wish to revise their habits or opinions. Therefore, it is a political process within a company to conclude that the purchasing process will, at least partly, use online auctions. This works best in companies with a centralized purchasing department. It may, therefore, be of interest to reorganize the purchasing structure, along with introducing electronic sourcing strategies. However, this should not be conducted in a "big bang" manner. Instead, a start in small online auction projects with specific auctionable goods seems to be the best strategy.

Occasionally, IT departments are one particular obstacle that we encounter. Very restrictive IT-departments are reluctant to allow their colleagues any Web-access. Such departments will not allow even the installation of Web browsers.

Another, more severe reason, is that IT-development departments tend to focus on overall solutions. IT-personnel usually favor strategic and expensive solutions. They would, for example, favor a complete restructuring of the procurement process, starting from a single pencil up to gigawatts of electrical power, through the same e-Procurement system. This kind of solution will be costly and will require a longer project for its implementation and, therefore, have no immediate effect. For some e-Procurement strategies, it is still unclear whether there will be a benefit in the long run, at all.

Online auctions, instead, have proven their immediate effect. It is possible to reduce the price paid for commodities, as well as for strategic materials and goods, within hours, after only a few days of preparation. Most importantly, online auctions tackle the major part of the total cost of ownership, namely, material costs. Standard e-Procurement techniques focus more on internal process costs, merely 7% of the total costs. Figure 4 shows the separation of the TCO into material and internal process costs.

## Identifying auctionable goods

Surprisingly, many purchasing departments do not have a detailed overview of the materials and goods they buy. Of course, they have the numbers available, which tell what percentage of the budget is spent on each good. However, to have a competitive advantage (starting with the purchasing process) means to know the structure of potential suppliers. How many competitors exist globally? What is the competitive market price? Does one or do few companies have the monopoly on selling a particular material? How easy is it for the company to switch to a competing material? Must all the material (or product) be from the same supplier? How reliable is the supplier? Is it useful to split strategic goods among several suppliers?

Purchasers often have no quantitative comparison available that would allow an easy identification of the most promising auctionable goods.

Therefore, it is often useful to analyze the material structure before identifying online auctions. The technique of portfolio analysis proved useful.

## How to set up an auction

Sometimes, it is very simple to set up an auction. In a reverse auction, the buyer usually knows quite well what he wants to buy. Therefore, a number of potential suppliers have to be identified and invited to participate in the auction. Apart from the current supplier, a number of potential suppliers usually is well-known already. Additionally and, if desired, a web-search can help to identify new suppliers. The online auction contract is settled. After that, the auction is conducted, and the purchasing contract is closed.

When auctioning strategic goods, however, the situation often is more complicated. For example, if the buyer does not want to depend on one supplier only, he can split the material into several slots, with the side condition that each supplier will get

*Figure 4: Separation of total costs into material and process costs*

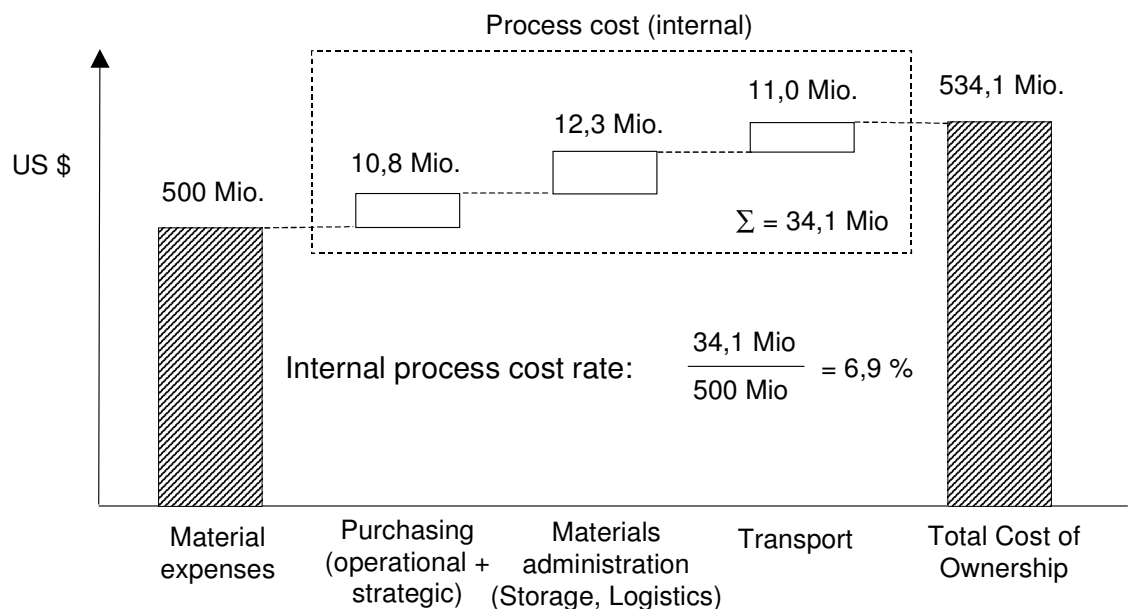

Source: Forrester Research

a contract for only one slot. Splitting material in slots is also of interest when the points of delivery are partitioned among time or space (e.g., in different countries).

To increase the competition, it is also possible to run two auctions on competing goods simultaneously, and define the contract in such a way that only the most competitive goods will be bought in the end. For example, if one material is of better quality but also more expensive, then the ratio of which goods to buy to what extent can depend on their price differences.

The concept of target price can be used to decrease barriers to the buyer conducting an online auction. The buyer is allowed to define a target price, which must be hit so that the buyer is obliged to sign the purchasing contract. If the target price is not hit, the buyer is free to sign. The definition of a good target price is not an easy task. In some occasions, and some markets, the definition of a competing target price had very good results. In other markets, the target price was missed completely. The problem occurred when the target price was set too competitively, which prevented the suppliers from starting to bid seriously at all. The most impressive results have been achieved by talking to the potential suppliers before the auction was conducted, and determining how they react to a more or less ambitious target price.

In most auctions, the bidders did not know each other, but there are markets where bidders can guess the other competitors easily. To decrease the risk of making special arrangements among bidders, it is useful to identify at least one or two additional bidders in the emerging global market. In addition, bidders can see, during the auction, how many competitors are online. Another strategy is to reveal the anonymity of the bidders with the start of the auction. To prevent arrangements between bidders, the auction is then conducted in 10-15 minutes only. This helps, for example, in markets where the competitors know and dislike each other.

The final problem encountered was the onlooker problem. Some suppliers were not interested in committing bids, but wanted to get an overview of the market prices of their competitors. In the markets where this problem can occur, multi-phase auctions are particularly useful. For example, the first phase of the auction is run as a normal auction, whereby a. first target price must be hit, before the participant is admitted to the second phase. In the second phase, the target price is usually lower, but not visible to onlookers, who are no longer allowed to cast bids. Two or three phases usually are enough.

To summarize, based on our experiences, different markets need customized variants of auctions. Some kinds of auctions are pretty straightforward. In other situations, markets are less competitive and, therefore, need additional techniques to ensure the finding of a fair price in an online auction.

# SUPPLIER REACTIONS

The Emporias' auction engine was able to reduce the costs for certain materials up to 46%. This is a great gain for the purchasers, but also a loss of profits for suppliers. Therefore, suppliers have started to realign their purchasing structures. Marketplaces that mainly focus on establishing contact between buyers and suppliers are not as critical. Online auctions are the critical part of the new E-business economy. They drastically increase competition in the global market place. The supplier problems can be narrowed to the following three questions:

- How can I be prevented from participating in an online auction?
- What is the best strategy when I am forced to participate in an online auction?
- How do I deal with the consequences arising from online auctions?

Suppliers have a great deal of possible actions at hand, ranging from strategic, long-lasting contracts with their purchasers to dramatic cost reductions by using online auctions in their own purchasing processes. Depending on the goods, the market peculiarities, and the geographic and contractual situations of the supplier, a number of additional actions can be taken to deal with that new situation. Aggressive suppliers do not fear participating in online market places, but try to use them as a chance to establish contact with new purchasers. As always: the early bird catches the worm.

To optimize suppliers' strategies in various markets, it is useful to conduct workshops with the goal of realigning organizational structures, defining reaction variants for each type of material, redefining pricing strategies and, finally, leading to an improved composition of the companies sales portfolio. See Figure 5 for a structuring of such a workshop.

Although it is possible to define general strategies for suppliers that have to react in online auctions, the results and, therefore, the behavior of competing suppliers greatly depends on the auction format. Therefore, auction formats have to be taken into consideration when defining the auction's individual bidding strategy. There are many possible choices of action, actions such as preventing an online auction, or breaking up a buying syndicate, but today, it seems, one of the most promising strategies for a supplier is to reduce material costs himself in online auctions.

# SUMMARY

In this chapter, we have introduced a number of auction formats and their impact on the conducting of competitive online auctions in business-to-business trading. Online auctions are among the Internet techniques that will become inevitable in the B2B purchasing area. They are rather easy to implement, and have a high, and immediate, gain of results in return.

*Figure 5: Workshop structure for suppliers' strategic realignment*

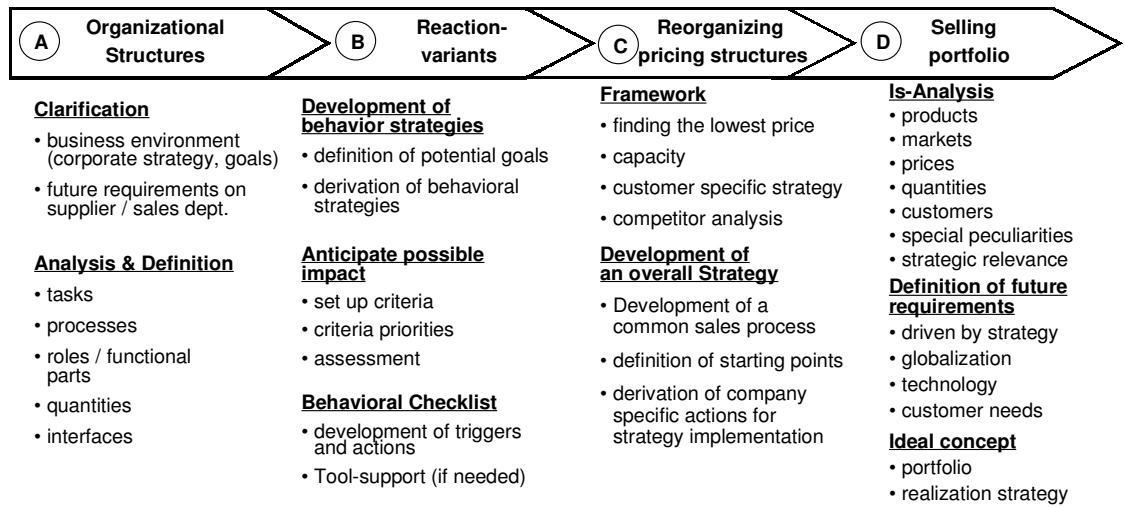

Current numbers and forecasts indicate the B2B E-business will explode even more in the coming years. However, the techniques that the new economy offers have to face the traditional fears and problems that the old economy — and the people in that economy — are still holding. Online auctions are one of the electronic sourcing techniques that will find their terra firma in the E-sourcing portfolio. However, the way to arrive there is not as easy and simple as the people had in mind in the early days of E-commerce.

# ACKNOWLEDGEMENTS

This work was supported by the Bayerisches Staatsministerium für Wissenschaft, Forschung und Kunst through the Bavarian Habilitation Fellowship and the German Bundesministerium für Bildung und Forschung through the Virtual Softwaereengineering Competence Center (ViSEK).